\documentclass[aip, amsmath, amssymb, reprint]{revtex4-1}

\usepackage{graphicx}
\usepackage{dcolumn}
\usepackage{bm}

\begin{document}

\title{Anomalous Hall effect in the Co-based Heusler compounds Co$_{2}$FeSi and Co$_{2}$FeAl}

\author{I.-M.\ Imort}
\email{iimort@physik.uni-bielefeld.de}
\homepage{www.spinelectronics.de}
\author{P.\ Thomas}
\author{G.\ Reiss}
\author{A.\ Thomas}
\affiliation{Thin Films and Physics of Nanostructures, Bielefeld University, Germany}

\begin{abstract}
The anomalous Hall effect (AHE) in the Heusler compounds Co$_{2}$FeSi and Co$_{2}$FeAl is studied in dependence of the annealing temperature to achieve a general comprehension of its origin. We have demonstrated that the crystal quality affected by annealing processes is a significant control parameter to tune the electrical resistivity $\rho_{xx}$ as well as the anomalous Hall resistivity $\rho_{ahe}$. Analyzing the scaling behavior of $\rho_{ahe}$ in terms of $\rho_{xx}$ points to a temperature-dependent skew scattering as the dominant mechanism in both Heusler compounds. 
\end{abstract}
\maketitle
\section{Introduction}
The Hall resistivity in ferromagnetic materials, $\rho_{xy}$ receives an extra contribution from the spontaneous magnetization M, which empirically results in the general formula $\rho_{xy}=R_{o}\mu_{0}H + R_{s}\mu_{0}M$, where $\mu_{0}H$ is the applied magnetic field. The coefficients $R_{o}$ and $R_{s}$ are characterized by the strength of the ordinary and anomalous Hall resistivity $\rho_{ohe}$ and $\rho_{ahe}$.  While the ordinary Hall effect\cite{Hall2} (OHE) is classically explained by the Lorentz force deflecting the moving charge carriers, the anomalous Hall effect\cite{Hall1} (AHE) has been a fundamental but controversially discussed aspect in solid-state physics since the 1950s. In several theories, both extrinsic and intrinsic mechanisms\cite{Karplus} have been proposed to be responsible for the AHE. The extrinsic origin of the AHE based on the skew scattering\cite{Luttinger, Smit} and side jump mechanisms\cite{Berger} has been ascribed to the asymmetric scattering of spin-polarized charge carriers in the presence of spin-orbit interaction\cite{Fert1, Fert2} whereas the intrinsic origin is closely associated with Berry phase effects of Bloch electrons.\cite{Karplus, Luttinger, Smit, Berger, Berry} 
In ferromagnetic materials a correlation between the anomalous Hall resistivity $\rho_{ahe}$ and the electrical resistivity $\rho_{xx}$ of the form $\rho_{ahe}\propto\rho_{xx}^{n}$ has been established, where n depends on the dominant origin of the AHE in a given material: $n=2$ for an intrinsic mechanism\cite{Karplus} 
 whereas $n=1$ or $n=2$ for the skew scattering\cite{Luttinger, Smit} or the side jump mechanisms\cite{Berger}, respectively. This correlation makes it possible to identify the predominant scattering mechanism. The superposition of all three contributions results in
 \begin{equation}
 \rho_{ahe}=a\rho_{xx} + b\rho_{xx}^2 
 \end{equation}
 The coefficient $a$ gains information about the skew scattering, while the coefficient $b$ corresponds to both the side-jump and the intrinsic mechanisms.    
Due to the importance of the spin-degree of freedom in spintronic devices such as magnetic sensors or magnetic memories, 
the capability of the AHE to generate and control spin polarized currents has lead to intense research on the AHE in new materials. Prominent candidates to be integrated in spintronic devices are Co-based Heusler compounds\cite{Balke} due to their predicted half-metallic behavior, i.e. 100\ \% spin-polarization at the Fermi level, and the high Curie temperatures. 
In the present work, the influence of crystallographic defects and atomic disorder on the AHE in the compounds Co$_{2}$FeSi and Co$_{2}$FeAl is studied. 
\section{Experimental}
For the preparation of all layer stacks conventional dc/rf magnetron sputtering at room-temperature is used in a high vacuum system with a base pressure of 1.0$\times$10$^{-7}$\ mbar and with a process pressure of 1.5$\times$10$^{-3}$\ mbar of Ar as sputtering gas. All films were deposited on a 5\ nm thick MgO film which was deposited on the MgO (001) substrate to improve the surface quality. Epitaxial thin films of Co$_{\rm2}$FeSi (CFS) and Co$_{\rm2}$FeAl (CFA) with a thickness of 20\ nm were deposited from pure (99,95\%) stoichiometric composition targets (Co 50\%, Fe 25\%, Si(Al) 25\%). Finally, they were protected from ambient oxidation by a 1.8\ nm thick MgO cover layer. After deposition, several lacker stacks of both Heusler compounds were ex-situ annealed at temperatures in the range from 27$^{\circ}$C (as-prepared) up to 700$^{\circ}$C in order to initiate ordering and to manipulate the amount of defects. Their crystalline structure has been determined by x-ray diffraction ($\theta-2\theta$) scans. Using photolithographic techniques and Ar-ion beam etching the layer stacks were patterned into 80-$\mu$m-wide and 200-$\mu$m-long Hall bar structures in order to measure the electrical and transverse Hall resistivity simultaneously. Magnetotransport measurements were performed in the temperature range from 300 K down to 3 K and in a magnetic field of up to 4 T. 
\section{Results and discussion}
In Fig. 1(a) the XRD patterns of the layer stacks are presented for different annealing temperatures. The appearance of the (002) and (004) diffraction peaks in the patterns reveals the B2-type structure in Co$_{2}$FeSi as well as in Co$_{2}$FeAl. 
Providing information about the degree of atomic disorder, the intensity ratio I$_{(002)}$/I$_{(004)}$ of the (002) and (004) diffraction peaks increases with rising annealing temperature in both Heusler compounds [see Fig. 1(b)]. This observation corresponds to a monotonous increase in crystallographic order on the Co-site of Co$_{2}$FeSi or Co$_{2}$FeAl in the temperature range from 300$^{\circ}$C to 700$^{\circ}$C. The structural refinement is additionally proven by the width of the (002) diffraction peak $\Delta(2\theta_{(002)})$ that becomes narrower with increasing annealing temperature and reaches a value of 0.55$^{\circ}$ for Co$_{2}$FeSi and 0.51$^{\circ}$ for Co$_{2}$FeAl at 700$^{\circ}$C. \cite{Ksenofontov}\\
\begin{figure}
\includegraphics{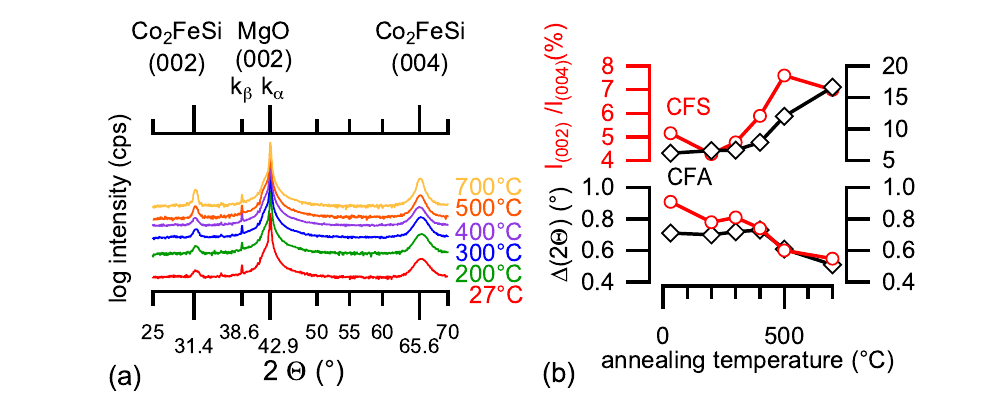}
\caption{\label{fig:pdf} (a) XRD ($\theta-2\theta$) scans of CFS at different annealing temperatures. (b) Intensity ratio $I_{(002)}/I_{(004)}$ and width of the (002) diffraction peak $\Delta(2\theta_{(002)})$ as a function of annealing temperature for both CFS (red circle) and CFA (black square).}
\end{figure}
In according to Matthiessen's rule, the electrical resistivity $\rho_{xx}(T)$ displayed in Fig. 2(a) can be separated into a temperature-independent resistivity $\rho_{xx0}$ below 20 K and a temperature-dependent resistivity $\rho_{xxT}(T)$, defined by
\begin{equation}
\rho_{xx}(T)=\rho_{xx0}+\rho_{xxT}(T)
\end{equation}
The residual resistivity $\rho_{xx0}$ is caused by the scattering at impurities and atomic disorder, while the scattering on lattice vibrations (phonons) and spin disorder (magnons) dominate $\rho_{xxT}(T)$ at higher temperatures. 
Above 100 K the resistivity curves show the characteristic weak and linear temperature dependence of Co-based Heusler compounds. \cite{Jakob}
With increasing annealing temperature a strong decrease of the resistivity can be observed except for the 200$^{\circ}$ C annealed layer stack. For Co$_{2}$FeAl the values of the resistivity show a similar tendency. As it can be seen in Fig. 2(b) an increase of defect concentration can be directly derived from the enhancement of the residual resistivity $\rho_{xx0}$ for layer stacks annealed at temperatures smaller than 500$^{\circ}$C (Co$_{2}$FeSi) or 700$^{\circ}$C (Co$_{2}$FeAl). 
The residual resistivity ratio $\rm{RRR}=(\rho_{xx}(300K)/\rho_{xx}(3K))$ as a qualitative tool to identify the degree of atomic disorder and lattice defect is displayed in Fig.2 as function of the annealing temperature for both Co$_{2}$FeSi and Co$_{2}$FeAl. 
The RRR starts to increase above an annealing temperature of 200$^{\circ}$C (Co$_{2}$FeSi) or 300$^{\circ}$C (Co$_{2}$FeAl). For Co$_{2}$FeSi the RRR reaches a maximum of 1.3 at 500$^{\circ}$C and decreases again for larger annealing temperature. The values of RRR for both compounds are comparable to literature values in the range of 1.2 to 1.4 for sputtered Heusler compounds.\cite{Raphael, Geiersbach} The enhancement of RRR with the annealing temperature can be explained by an improvement of crystallographic order and, hence, a reduction of defect concentration in the Heusler lattice. The differences in the RRR of both compounds [Fig. 2(b)] hints to a larger defect concentration in Co$_{2}$FeAl as compared with Co$_{2}$FeSi. \\
\begin{figure}
\includegraphics{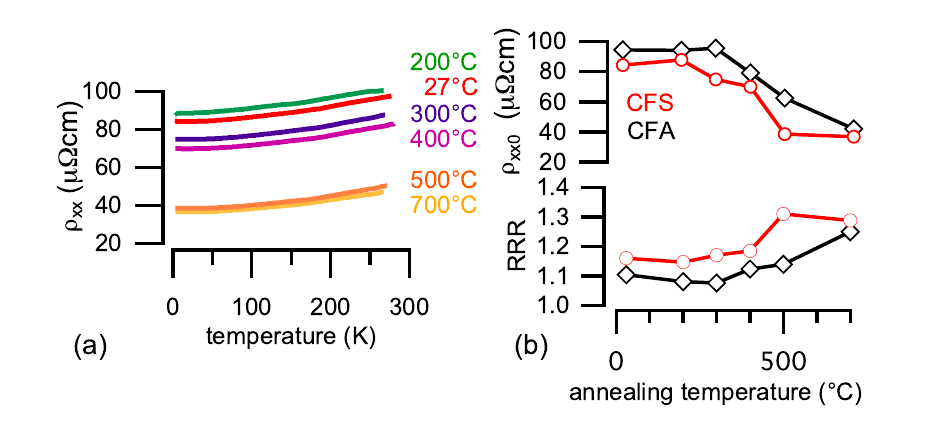}
\caption{\label{fig:pdf}(a) Electrical resistivity $\rho_{xx}(T)$ for CFS films annealed at different temperatures. (b) Residual resistivity $\rho_{xx0}$ and RRR values as function of annealing temperature for CFS (red circle) and CFA (black squares).}
\end{figure}
 \begin{figure}
\includegraphics{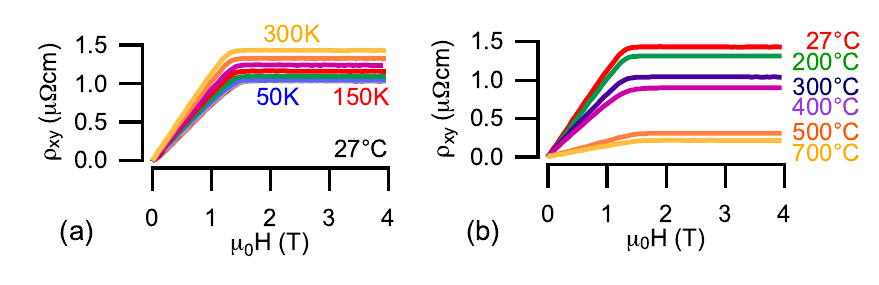}
\caption{\label{fig:pdfartt} a) $\rho_{xy}(\mu_{0}H)$ at selected temperatures is shown for the as-prepared (27$^{\circ}$C) CFS layer stack. (b) Annealing temperature dependence of the room-temperature $\rho_{xy}(\mu_{0}H)$ of CFS layer stacks.}
\end{figure}
In Fig. 3(a) the magnetic field dependence of the Hall resistivity $\rho_{xy}(\mu_{0}H)$ is shown for the as-prepared (27$^{\circ}$C) layer stack at several temperatures. The values of  $\rho_{ahe}$ corresponding to each temperature can be extracted by extrapolation of the $\rho_{xy}(\mu_{0}\rm{H})$ data curves taken from the high back to zero field. As it is shown in Fig. 3 the values of $\rho_{ahe}$ reveal only a weak temperature dependence. Conversely, $\rho_{ahe}$ measured at room-temperature strongly changes with the annealing temperature from 1.436 $\mu\Omega$cm at 27$^{\circ}$C down to 0.219 $\mu\Omega$cm at 700$^{\circ}$ C.
\begin{figure}
\includegraphics{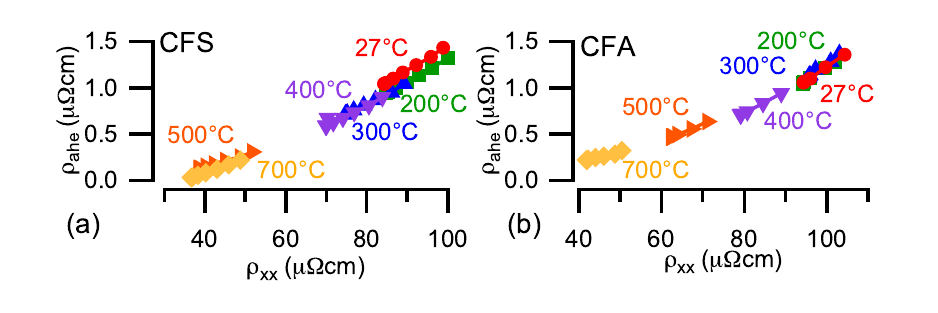}
\caption{\label{fig:epsart}$\rho_{ahe}(T)$ vs. $\rho_{xx}(T)$ for different annealing temperatures for CFS (a) and CFA (b).}
\end{figure}  
Fig. 4 represents $\rho_{ahe}(T)$ versus $\rho_{xx}(T)$ for different annealing temperatures in the range from 27$^{\circ}$C up to 700$^{\circ}$C. Each point corresponds to one measuring temperature for Co$_{2}$FeSi (a) and Co$_{2}$FeAl (b). 

The obvious correlation between the crystal quality characterized by $\rho_{xx}(T)$ and $\rho_{ahe}(T)$ suggests the presence of scattering mechanisms in both resistivities caused by defect concentration and atomic disorder. Therefore, the scaling behavior between $\rho_{ahe}(T)$ and $\rho_{xx}(T)$ can be investigated in order to provide information about the dominant scattering mechanism responsible for the appearance of the underlying anomalous Hall effect.\cite{Tian}\\
The following discussion is based on the data analysis by Vidal et al.\cite{Vidal, Gerber_1, Gerber_2} Taking into account the scaling behavior of both resistivities [see Fig. 4], the effect of different scattering mechanisms is analyzed. The anomalous Hall resistivity $\rho_{ahe}(T)$ reveals an almost identical temperature dependence as the electrical resistivity $\rho_{xx}(T)$. According to the temperature-dependent separation of $\rho_{xx}(T)$ [see Eq.(2)], the anomalous Hall resistivity $\rho_{ahe}(T)$ defined in Eq.(1) can be divided in an analogous way
\begin{equation}
\rho_{ahe}=\rho_{ahe,0} + (a+2b\rho_{xx0})\rho_{xxT} + b\rho_{xxT}^2
\end{equation}
where $\rho_{ahe,0}=(a\rho_{xx0}+b\rho_{xx0}^2)$ describes the residual anomalous Hall resistivity.
Below 20 K both $\rho_{xx}$ and $\rho_{ahe}$ are approximately constant, while above both increase. This observation enables a detailled analysis of the temperature dependence by subtraction of the temperature-dependent (T$_{h}$) and temperature-independent (T$_{0}$) data, where the T$_{0}$ data corresponds to the data at 20 K. This involves the following expression
\begin{equation}
\Delta\rho_{ahe}=(a+2b\rho_{xx0})\Delta\rho_{xxT}+b\Delta\rho_{xxT}^2
\end{equation}
To simplify Eq.(4), the rescaled electrical and anomalous Hall resistivity are defined by $\Delta\rho_{xxT}=\rho_{xxT}(T_{h})-\rho_{xxT}(T_{0})$ and $\Delta\rho_{ahe}=\rho_{ahe}(T_{h})-\rho_{ahe}(T_{0})$. 
In Fig. 5 the rescaled quantities are plotted for various annealing temperatures.
\begin{figure}
\includegraphics{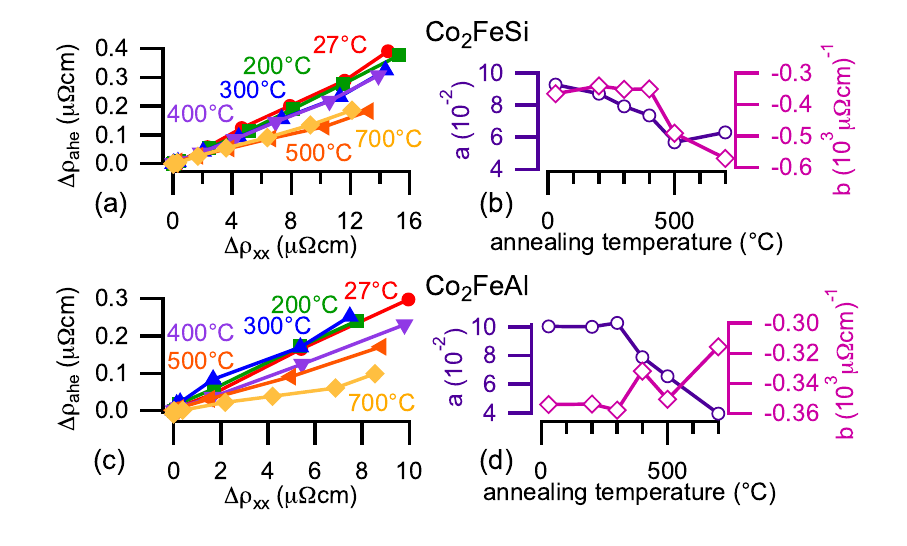}
\caption{\label{fig:epsart} Left column: $\Delta\rho_{ahe}$ vs. $\Delta\rho_{xx}$ for CFS (a) and CFA (c). Right column: Linear parameter a (purple squares) and quadratic parameter b (pink circles) against annealing temperature for CFS (b) and CFA (d). }
\end{figure}
Within the experimental errors, all rescaled data curves for Co$_{2}$FeSi have the same slope except the data curves for 500$^{\circ}$C and 700$^{\circ}$C annealing temperature. The common slope of these two data curves is marginally smaller compared to those of the other annealing temperatures. For Co$_{2}$FeSi both parameters $a$ and $b$ obtained by fitting of Eq.(4) to the data monotonously decrease with increasing annealing temperature. In contrast to this, for Co$_{2}$FeAl the data for annealing temperatures of 27$^{\circ}$C, 200$^{\circ}$C and 300$^{\circ}$C coincide, while the slope of the curves for larger annealing temperatures decreases.
Above 300$^{\circ}$C the scattering parameter $a$ increases with annealing temperature whereas the parameter $b$ decreases. 
The rescaled data curves for both Co$_{2}$FeSi and Co$_{2}$FeAl exhibit an approximately linear relation between $\Delta\rho_{ahe}$ and $\Delta\rho_{xx}$. As a result, the skew scattering mechanism can be supposed to be dominant in both Heusler compounds. Below an annealing temperature of 300$^{\circ}$C in Co$_{2}$FeAl or 400$^{\circ}$C in Co$_{2}$FeSi the skew scattering mechanism appears to be weakly affected by the temperature [see Fig. 5(b) and 5(d)]. From the fitting procedure, it is clearly visible that the values of the scattering parameter $b$ are of the order of 10$^{-4}(\mu\Omega cm)^{-1}$. Taking into account the residual resistivity $\rho_{xx0}$ of the order 10$^{1}(\mu\Omega cm)$, the linear coefficient in Eq.(4) $(a+2b\rho_{xx0})\approx 10^{-2}$ is dominated by the skew scattering parameter $a$.
The temperature dependence of the skew scattering, however, indicates that the scattering centers are not mainly crystal defects, because this would lead to a temperature independent contribution. In contrast, scattering at phonons and magnons is strongly temperature dependent. We thus conclude, that scattering at magnons is responsible for the observed temperature dependent $\rho_{ahe}(T)$.
\nocite{Zeng, Venka, Sangiao, Koetzler}
\section{Summary}
In conclusion, we have performed a systematic study of the AHE in the Co-based Heusler compounds Co$_{2}$FeSi and Co$_{2}$FeAl. The crystallographic quality varied by the annealing temperature reveals to be a convenient control parameter tuning the electrical as well as the anomalous Hall resistivity $\rho_{xx}(T)$ and $\rho_{ahe}(T)$ in both Heusler compounds. Based on this idea,  skew scattering at magnons could be identified as main reason for the temperature dependent anomalous Hall effect in Co$_{2}$FeSi and Co$_{2}$FeAl.
\section*{Acknowledgments}
A.T. and I.-M. I. are supported by the NRW MIWF. G.R. and I.-M.I  were supported by the DFG priority program SpinCaT.
\section{Refernces}
\providecommand{\noopsort}[1]{}\providecommand{\singleletter}[1]{#1}%

\end{document}